\documentclass{emulateapj}
\shorttitle{Spectra of Debris Disks}
\shortauthors{Dodson-Robinson et al.}
\usepackage{multirow}
\usepackage{lscape}
\usepackage[left=1.25in,right=0.75in,top=1.75in,bottom=0.25in,paperwidth=8.5in,paperheight=11in]{geometry}
\begin{document}

\title{A {\it Spitzer} IRS Study of Debris Disks around
Planet-Host Stars}

\author{Sarah E. Dodson-Robinson\altaffilmark{1,5},
C. A. Beichman\altaffilmark{2},
John M. Carpenter\altaffilmark{3},
Geoffrey Bryden\altaffilmark{4}}

\altaffiltext{1}{Astronomy Department, University of Texas, 1 University
Station C1400, Austin, TX 78712, USA; sdr@astro.as.utexas.edu}

\altaffiltext{2}{NASA Exoplanet Science Institute, California Institute
of Technology, 770 S. Wilson Ave, Pasadena, CA 91105, USA}

\altaffiltext{3}{Department of Astronomy, California Institute of
Technology, Mail Code 249-17, 1200 E. California Blvd, Pasasdena, CA
91105, USA}

\altaffiltext{4}{Jet Propulsion Laboratory, California Institute of Technology, 4800 Oak Grove Dr, Pasadena, CA 91109, USA}

\altaffiltext{5}{Former Spitzer Fellow}

\begin{abstract}

Since giant planets scatter planetesimals within a few tidal radii of
their orbits, the locations of existing planetesimal belts indicate
regions where giant planet formation failed in bygone protostellar
disks. Infrared observations of circumstellar dust produced by colliding
planetesimals are therefore powerful probes of the formation histories
of known planets. Here we present new {\it Spitzer} IRS
spectrophotometry of 111 Solar-type stars, including 105 planet hosts.
Our observations reveal 11 debris disks, including two previously
undetected debris disks orbiting HD~108874 and HD~130322.  Combining our
32~$\mu$m spectrophotometry with previously published MIPS photometry,
we find that the majority of debris disks around solar-type stars have
temperatures in the range $60 \la T_{\rm dust} \la 100$~K.

Assuming a dust temperature $T_{\rm dust} = 70$~K, which is
representative of the nine debris disks detected by both IRS and MIPS,
debris rings surrounding Sunlike stars orbit between 15 and 240~AU
depending on the mean particle size. Our observations imply that the
planets detected by radial-velocity searches formed within 240~AU of
their parent stars. If any of the debris disks studied here have mostly
large, blackbody emitting grains, their companion giant planets must
have formed in a narrow region between the ice line and 15~AU.

\end{abstract}


\section{Introduction}
\label{intro}

Sometimes the giant planets that failed to form reveal as much about the
birth environments of other Solar Systems as the giant planets we
actually find. Our objective is to characterize circumstellar dust
produced by colliding planetesimals orbiting planet hosts and use the
orbital radii of such planetesimals to constrain the locations where the
known giant planets may have formed.  Radial-velocity searches do not
yet have the time baseline to detect giant planets more than $\sim 7$~AU
from their host stars. Nevertheless, we expect that giant planets do not
reside within observed debris disks---which typically orbit at 10~AU or
more---because giant planets scatter planetesimals within a few Roche
radii of their orbits (e.g.  Bryden et al.\ [2000], Thommes et al.\
[2003], Ida \& Lin [2004], Goldreich et al.\ [2004]). As long as
planetesimal belts don't migrate on a large scale, grain temperature
measurements that constrain the location of circumstellar dust highlight
the regions in long-gone protostellar disks where giant planet formation
didn't succeed.

The ``Vega Phenomenon''---an excess of infrared emission caused by
circumstellar dust---was first observed by Aumann et al.\ (1984) using
the {\it Infrared Astronomical Satellite}.  Since then, observations by
the {\it Spitzer} Space Telescope have revealed the temperatures and
locations of scores of extrasolar Kuiper Belts (e.g.\ Chen et al.\
[2006], Hillenbrand et al.\ [2008]).  Here we combine new {\it Spitzer}
IRS spectrophotometry of planet hosts with previously published MIPS
data, allowing us to constrain grain temperatures and planetesimal belt
locations and exclude regions inhabited by circumstellar debris from the
possible giant planet formation zone.

Massive planets with small orbital radii almost certainly formed at
larger radii, where they had access to abundant solid material
(Bodenheimer et al.\ 2000), and migrated inwards through either
dynamical interactions with massive gas disks (Lin et al.\ 1996) or
Kozai cycles triggered by a companion (Libert \& Tsiganis 2009).
Migration will disrupt any planetesimal belt in the giant planet's path,
increase the collision rate of planetesimals, and accelerate the
evolution of the planetesimal belt to reduce the debris production at
older ages. Indeed, migration of the gas-giant planets in our own Solar
System and the subsequent disruption of the Kuiper Belt is one possible
explanation for the Late Heavy Bombardment that occurred in the inner
solar system 3.9 Gyr ago (Gomes et al.\ 2005). Raymond et al.\ (2006)
found that if migration occurs quickly and the planet settles at an
orbital radius $< 0.25$~AU, the planetesimal disk can regenerate to form
terrestrial planets, and consequently may form a long-lived debris disk.
However, disk regeneration is suppressed if the migration stops at a
radius $a > 0.5$~AU and the debris may dissipate rapidly.

The evidence for brighter debris disks around planet hosts than
non-hosts has at times been intriguing (e.g.\ Beichman et al.\ 2005,
K\'{o}sp\'{a}l et al.\ 2009), but most studies show no statistically
significant differences between the circumstellar dust populations of
planet hosts and other stars (Moro-Mart\'{i}n et al.\ [2007], Bryden et
al.\ [2009]). Despite a few compelling exceptions---for example, the
existence and location of Fomalhaut b was predicted through analysis of
the shape and surface brightness profile of Fomalhaut's debris disk
(Quillen [2006], Kalas et al.\ [2008])---statistical analysis indicates
that known giant planets tend to be dynamically decoupled from any
planetesimal belts orbiting their host stars.


Since 60\% of the exoplanets discovered at the time of this writing have
semimajor axes $> 0.5$~AU\footnote{Based on histograms provided by the
Exoplanet Encyclopedia, http://exoplanet.eu}, we expect that few
planetesimal belts in the path of migrating planets were able to
regenerate according to the Raymond et al.\ (2006) scenario. {\it The
simplest hypothesis, then, is that migratory planets formed nearer their
host stars than the dust-producing planetesimal belts generally studied
with IRAS and Spitzer}. Migrating inward, they left the planetesimal
belts intact. Since giant planet formation efficiency decreases with
distance from the star (Safronov [1969], Pollack et al.\ [1996]), we
expect that today's planetesimal belts had too little solid mass to
overcome the growth slowdown inherited from their wide orbits.
Excepting cases where observations conclusively demonstate that a
particular debris disks and giant planet dynamically interact (Quillen
[2006], Lovis et al.\ [2006], Kalas et al.\ [2008]), the ensemble of
debris disks orbiting known radial-velocity planet hosts provides an
upper limit to the semi-major axis of the locations where most of the
known planets could have formed.



In \S \ref{sample} we describe our sample of target stars. In \S
\ref{observations} and \S \ref{detections}, we detail our procedure for
reducing {\it Spitzer} IRS data and detecting infrared excesses.  In \S
\ref{debrisdisks} we discuss our debris disk detections. We focus
specifically on the luminosity and location of the dust in \S
\ref{dtemp}.  Finally, we list our conclusions in \S \ref{conclusions}.

\section{Sample Selection and Observations}
\label{sample}

The main sample of extrasolar planetary systems was selected for {\it
Spitzer} program 40096 (PI: J. Carpenter) from the compilation reported
in the Extrasolar Planets Encyclopaedia (http://exoplanet.eu) as of
February 16, 2007. At the time, there were 182 known extrasolar systems
containing 212 planets. From this parent sample, we selected extrasolar
planetary systems that met the following criteria:

\begin{enumerate}

\item The planets were discovered by radial velocity techniques, or from
transiting surveys and later confirmed by radial velocity measurements.
We excluded planets identified from microlensing surveys, pulsar
timing experiments, and direct imaging of star forming regions.

\item Target stars had distances available from Hipparcos parallaxes. 

\item Targets had either main-sequence or sub-dwarf luminosity class to
focus the survey of solar-analogs.

\item Targets were bright enough to reach a signal to noise ratio of 20
in a spectrophotometric bandpass between 30-34~$\mu$m within 20 AOR
cycles of 120 seconds.

\end{enumerate}

These criteria narrowed the list to 143 planet hosts, all within a
distance of 100 pc, containing 171 planets. Stellar age estimates for
the sample are available from the detailed spectroscopic work by Valenti
\& Fischer (2005), and/or Ca II H and K stellar activity measurements
(Saffe et al.\ [2005], Wright et al.\ [2004]). We cross-correlated this
list of stars with the Reserve Object Catalog using a $5 \times 5$
arcminute search area and identified 34 extrasolar planet hosts with
previous IRS observations, leaving 109 stars for our IRS observing
program. We excluded seven stars observed in program 40096 from our
final sample because of poor sky subtraction or stray light from nearby
sources.

After the observations were complete, the existence of the planet
orbiting HD~188753 (Konacki 2005) was challenged by Eggenberger et al.\
(2007). Here we present our observations of HD~188753 but classify it as
a non-host.  Information about our sample, including the number of
planets orbiting each star and the 70~$\mu$m dust luminosity, is given
in Table 1. Effective temperatures were taken from the following sources
in order of preference: (1) Valenti \& Fischer (2005, abbreviated VF05
in Table 1), (2) Santos et al.\ (2004b), and (3) the planet discovery
paper.

We also present observations of nine solar-type stars with previously
unpublished data from other IRS programs. From IRS program 41 (PI: G.\
Rieke), we included the planet hosts HD~50554, HD~52265, HD~117176
(70~Vir), and HD~134987, bringing the total number of planet hosts in
our sample to 105. Finally, we included the stars without planets
HD~166, HD~33262 and HD~33636 from program 41, and HD~105211 and
HD~219482 from program 2343 (PI: C.\ Beichman). Our final sample
includes 111 stars, six of which have no detected planets.

Of the stars in our sample, 109 are FGK dwarfs and two, GJ~876 and
GJ~581, are M~dwarfs. All but three of the stars have 70~$\mu$m
photometry either from the MIPS component of our observing campaign,
described by Bryden et al.\ (2009), or from previous surveys.  All but
six of our targets are older than 1~Gyr, which reduces any spurious
statistical effects that might result from the well-know correlation
between debris disk brightness and stellar age (Su et al.\ 2006). Thus,
our discovered debris disks contain information about the planetesimal
populations of mature planetary systems such as our own.

The Infrared Spectrograph ({\em IRS}) aboard the {\em Spitzer} Space
Telescope observed the 102 targets from program 40096 between July 2007
and January 2008 ({\em IRS} campaigns 42-47). For each star, we recorded
five six-second exposures in the SL1 module to characterize the stellar
photosphere between 7 and 14~$\mu$m. All stars received a minimum
exposure time of 30 seconds (five six-second cycles) in the LL2 and LL1
modules, with extra cycles and/or ramp time added as needed for faint
targets to detect the photosphere at the 2-3$\sigma$ level in each
resolution element. We imposed a maximum integration time of 120 sec
with 20 cycles to keep the AORs to a reasonable length. We also required
a minimum of five cycles to achieve the redundancy needed to identify
cosmic rays and bad pixels, and to compute uncertainties based on
repeatability of the spectral extraction.  The IRS follow-up
observations of 70~$\mu$m excesses from IRS programs 41 and 2343 went
deeper in the LL1 module, reaching $S/N \approx 20$ per resolution
element between 30 and 34~$\mu$m.

\section{Data Analysis}
\label{observations}

Bad pixels were identified using the campaign-specific rogue pixel masks
provided by the {\em Spitzer} Science Center. We used the IRSCLEAN
package to remove bad pixels from the {\em Spitzer} pipeline's basic
calibrated data.  Sky subtraction and spectral extraction were performed
with the SMART data reduction package developed by the {\em IRS}
instrument team at Cornell (Higdon et al.\ 2004). Each star was observed
at two positions along the slit, Nod 1 and Nod 2. For each nod, we
constructed sky and stray-light correction frames by taking the median
of all exposures performed in the same module but opposite nod.

To make the end-to-end spectrum, we clipped unreliable data points from
the beginning and end of each order. We retained SL1 data from
7--14$\mu$m, LL2 data from 14--20$\mu$m, and LL1 data from 20--35$\mu$m.
We imposed a requirement that the final spectrum be continuous over
order crossings to correct for the different slit losses in each order.
A smooth spectrum was produced by correcting the LL2 spectrum with a
constant scaling factor to match the SL1 flux in the overlap region at
14$\mu$m, then scaling LL1 fluxes to match LL2 in the 20$\mu$m overlap
region.

The final step in our data reduction process was to create a model of
the star's photosphere. For each FGK star, we used the Atlas9 grid of
model atmospheres for stars with solar abundance ratios (Kurucz 1992).
Following the recommendations of Bertone et al.\ (2004), we used the
NextGen model atmospheres (Hauschildt et al.\ 1999) for photospheres of
the M~Dwarfs GJ~876 and GJ~3021, which have effective temperatures less
than 4000~K.

Photosphere models were fitted using Hipparcos parallaxes and
Hipparcos/Tycho $BT$ and $VT$ magnitudes (Perryman et al.\ 1997; H{\o}g
et al.\ 2000) transformed to the Johnson photometric system, 2MASS $J$,
$H$ and $Ks$ magnitudes (Cutri et al.\ 2003), and the measured
temperature of each star from planet-search spectra. Sources of $T_{\rm
eff}$ measurements are listed in Table 1. When available, $R$ and $I$
photometry from Bessell (1990) with assumed $0.1$~mag uncertainties were
added to further constrain the photosphere models. Each model
photosphere was extrapolated to mid-infrared wavelengths using a simple
blackbody extension to the optical and near-infrared fit.

In the next section, we describe our method for detecting
non-photospheric emission, the signature of circumstellar dust.

\section{Detecting Excess Emission}
\label{detections}

To construct the infrared excess spectra, we subtracted the model
photosphere from the IRS spectrum of each star and calculated the
fractional flux in each resolution element relative to the photosphere:
\begin{equation}
f_{\nu} = {F_{\nu}({\rm observed}) - F_{\nu}({\rm photosphere}) \over
F_{\nu}({\rm photosphere})} .
\label{fracexcess}
\end{equation}
{\em Spitzer} MIPS 24~$\mu$m observations of a sample of FGK dwarfs with
median age 4~Gyr have demonstrated that only $\sim 1$\% of main-sequence
Solar-type stars have dust with luminosity $L_{\rm dust} / L_* >
10^{-5}$ at 24~$\mu$m (Bryden et al.\ 2006), which originates from
asteroid-belt analogs at 3-4~AU from Sunlike host stars. The fractional
excess spectrum of a solar-type star with a typical debris disk should
stay near zero until between 25 and 30~$\mu$m, at which point the dust
gives it a rising slope. With the low likelihood of detecting
short-wavelength excesses in mind, our first step in calculating the
fractional excess spectrum was to assume the first 10 resolution
elements (20.93-22.45~$\mu$m) of our clipped LL1 spectra defined the
photosphere, such that
\begin{equation}
\frac{1}{\Delta \nu} \int_{20.93 \mu {\rm
m}}^{22.45 \mu {\rm m}} f_{\nu} d \nu = 0,
\label{photosphere}
\end{equation}
where $\Delta \nu$ is the frequency interval between 20.93 and
22.45~$\mu$m.

In the previous section we described using the order overlap regions to
correct for differential slit losses, a procedure that produces a smooth
but not flux-calibrated spectrum. In order to perform spectrophotometry
of our sources it is essential that we avoid the unreliable parts of our
spectra---the well-known SL1 order curvature and the LL1 24~$\mu$m flux
deficit\footnote[1]{See a description of these problems at
http://ssc.spitzer.caltech.edu/irs/features.}.  Pinning the
short-wavelength end of LL1 to the photosphere model defines a flux
calibration using an instrumentally reliable part of our spectra that
should contain only photospheric emission. In two cases where we
did observe a rising fractional excess slope shortward of the
20.93~$\mu$m LL1 boundary (HD~166 and HD~219482), we used the first 10
resolution elements of LL2 (14.24-15.00~$\mu$m) to define the
photosphere.

Following the approach of Beichman et al.\ (2006b) and Lawler et al.\
(2009), we defined a filter spanning 30-34~$\mu$m which is primarily
sensitive to dust with $T \approx 115$~K. Such dust would be located
between 8 and 10~AU from a Sunlike star assuming blackbody grains. To
search for debris disks, we calculate the frequency-weighted mean excess
relative to the photosphere in the 30-34~$\mu$m filter. The excess
measurement error $\sigma$ based on uncertainties propagated through the
{\it Spitzer} pipeline and photospheric fitting error is
\begin{equation}
\sigma^2 = \frac{1}{\Delta \nu} \int_{30 \mu {\rm m}}^{34 \mu
{\rm m}} \sigma_{\nu, {\rm obs}}^2 d\nu + (0.02 F_*)^2 .
\label{sigma}
\end{equation}
In equation \ref{sigma}, $\sigma_{\nu, {\rm obs}}$ is the uncertainty in
each resolution element from the {\it Spitzer} pipeline propagated
through the SMART data reduction, $\nu$ is the frequency of each
resolution element, and $\Delta \nu$ is the frequency difference between
30 and 34~$\mu$m.  The star flux $F_*$ in equation \ref{sigma} is simply
the frequency-weighted average flux of the photosphere model between 30
and 34~$\mu$m:
\begin{equation}
F_* = \frac{1}{\Delta \nu} \int_{30 \mu{\rm m}}^{34 \mu{\rm m}}
F_{\nu}({\rm photosphere}) d \nu .
\label{fstar}
\end{equation}
Following Beichman et al.\ (2006b), we impose a minimum fractional
photosphere fitting uncertainty of 2\%, which gives the $(0.02 F_*)^2$
component of the variance. In writing equation \ref{sigma} we assume
that the observational errors in each resolution element are independent.

If the errors reported by the {\it Spitzer} pipeline/SMART and the 2\%
photosphere-fitting uncertainty account for all of the dispersion in our
infrared excess measurements, we expect only one source to have more
than a $2 \sigma$ flux deficit. However, a closer look at the
distribution of excesses in our sample reveals seven stars with
2-3~$\sigma$ flux deficits between 30 and 34~$\mu$m, indicating that
there is an extra source of error in our infrared excess measurements.
Before conclusively identifying any debris disks, we follow Beichman et
al.\ (2006b) and require that they have a significant infrared excess
relative to the dispersion of the entire sample, not just internal
measurement and fitting errors.  Accordingly, we rank each debris disk
candidate by the significance of its infrared excess against the
internal error $\sigma$.  With such a ranking system we can identify
low-luminosity debris disks in high signal-to-noise observations while
rejecting debris disk candidates that appear to be bright but also have
large measurement errors.

\begin{figure}
\plotone{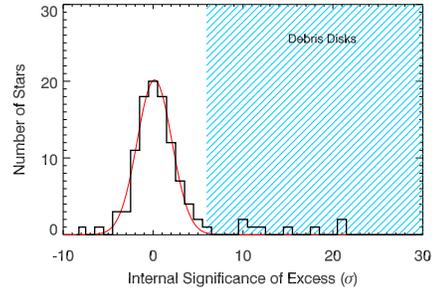}
\caption{Debris disks were identified by fitting a Gaussian
distribution, plotted in red, to a histogram of fractional 30-34~$\mu$m
excesses. The histogram variable, shown in black, is the number of
standard deviations $\sigma$ of the measured excess from a perfect
photosphere, where $\sigma$ is the frequency-weighted quadrature sum of
the internal measurement error and photosphere fitting error. Our sample
has an overall fractional excess dispersion of $\Sigma = 1.93 \sigma$
(see equation 5).  The blue cross-hatched region contains our debris
disks, which have fractional infrared excesses greater than $3 \Sigma$.
Our sample includes 11 stars with statistically significant 30-34~$\mu$m
excesses.}
\label{excess}
\end{figure}

We construct a histogram in the variable 
\begin{equation}
N = {F_{\rm obs} - F_* \over \sigma} ,
\label{nsigma}
\end{equation}
where $N$ is the number of standard deviations $\sigma$ from a perfect
photosphere in the 30-34~$\mu$m filter and $F_{\rm obs}$ is the
frequency-weighted average observed flux.  We can calculate $F_{\rm
obs}$ by replacing the predicted photospheric flux $F_{\nu} ({\rm
photosphere})$ in equation \ref{fstar} with the observed flux $F_{\nu}
({\rm observed})$.  Figure \ref{excess} shows the resulting histogram
along with its best-fit Gaussian function. Since a Gaussian function is
an excellent fit to the histogram in Figure \ref{excess}, the extra
error source is random rather than systematic. From the standard
deviation of our model Gaussian, we find that the dispersion of $N$
across our sample is $\Sigma = 1.93 \sigma$. Accordingly, we increase
our spectrophotometric error estimates by a factor of 1.93 to select
only debris disks with $3 \Sigma$ or greater excesses. The
cross-hatching in Figure \ref{excess} demarcates the part of the $N$
distribution where the debris disks lie.  With the aforementioned
detection procedure, we find statistically significant 32~$\mu$m
excesses in the spectra of 11 stars. In the next section we discuss the
spectra of our debris disks.

\section{Debris Disks}
\label{debrisdisks}

Table 2 shows the results of our 30-34~$\mu$m spectrophotometry.  The HD
numbers of debris disks appear in bold in the table. We detect two new
debris disks orbiting the planet hosts HD 108874 and HD 130322. We also
measure 30-34~$\mu$m emission from nine other debris disks previously
detected at 70~$\mu$m, four of which orbit planet hosts. The table also
gives fractional dust fluxes at 30-34~$\mu$m and total luminosities for
each star in the sample. Assuming the dust is a single-temperature
blackbody, its luminosity is
\begin{equation}
{L_{\rm dust} \over L_*} = {F_{\rm dust} \over F_*}
{kT_{\rm dust}^4 \left ( \exp \left [ h \nu / kT_{\rm dust} \right ] -1
\right ) \over h \nu T_*^3},
\label{lumdust}
\end{equation}
where $F_{\rm dust} = F_{\rm obs} - F_*$.  When the dust temperature is
not known, one can use equation \ref{lumdust} to calculate the minimum
dust luminosity, assuming the emission peak of the dust is centered in
the spectrophotometric filter.  For peak dust luminosity at 32~$\mu$m,
$T_{\rm dust} = 115$~K.

One can calculate the true luminosity ratio by detecting the debris disk
at multiple wavelengths. We write equation \ref{lumdust} once each for
our 32~$\mu$m spectrophotometric fluxes and the MIPS 70~$\mu$m values
from the literature and solve the resulting system for $T_{\rm dust}$
and $L_{\rm dust} / L_*$. Note that equation \ref{lumdust} assumes that
the dust has a single temperature---debris disks containing a large
range of grain sizes or occupying a wide range of distances from the
star will have a range of dust temperatures.

Table 2 contains three different types of measurements of $L_{\rm dust}
/ L_*$: true values, lower limits, and pseudo-upper limits. For stars
without any measured infrared excess, we use equation \ref{lumdust} and
set $F_{\rm dust} = 3 \Sigma$ to calculate the maximum dust luminosity
{\it assuming the excess emission comes from single-temperature dust at
115~K.} For cold, faint dust undetected by both MIPS and IRS, the true
luminosity may be higher than the maximum value we calculated if the
dust emission peaks at very long wavelengths. Lower limits to the dust
luminosity arise for debris disks with a detection at only one
wavelength and true values come from detections at multiple wavelengths
(see \S \ref{dtemp}).

\begin{figure}
\plotone{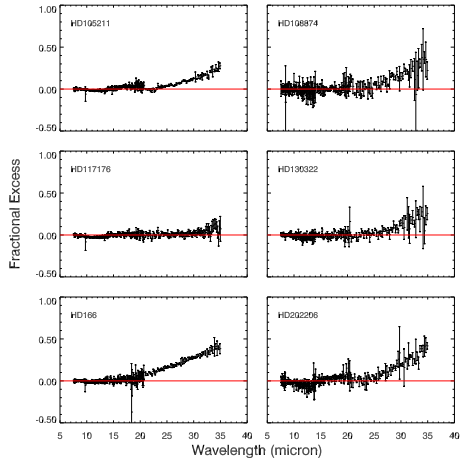}
\caption{The debris disks in our sample have fractional excess
spectra that rise toward longer IRS wavelengths. Only a few have any
noticeable rise near 24~$\mu$m, consistent with the paucity of 24~$\mu$m
excesses observed by the MIPS instrument. Several
spectra show the well-known SL1 order droop. HD~108874,
HD~117176, HD~130322 and HD~202206 are planet hosts.}
\label{debrisdisks1}
\end{figure}

\begin{figure}
\plotone{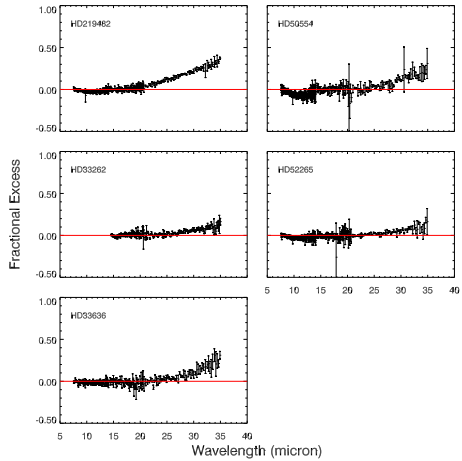}
\caption{Fractional excess spectra of detected debris disks, continued.
No SL data points are included for HD 33262 because the SL1 spectra were
partially saturated. HD~50554 and HD~52265 are planet hosts.}
\label{debrisdisks2}
\end{figure}

\begin{figure}
\epsscale{0.8}
\plotone{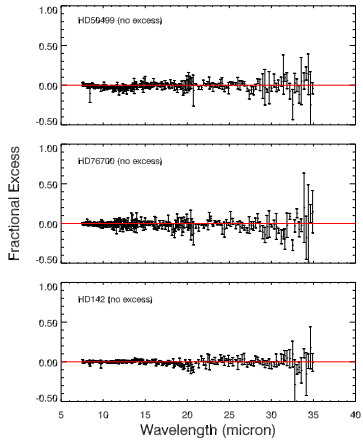}
\caption{Fractional excess spectra of stars with no significant
30-34~$\mu$m emission. These stars show no noticeable rise in excess
toward long wavelengths.}
\label{controls}
\end{figure}

Figures \ref{debrisdisks1} and \ref{debrisdisks2} show the spectra of
the debris disks in our sample. Each debris disk has little or no
fractional infrared excess in the IRS SL1 and LL2 modules but a positive
fractional excess slope in the long wavelengths of the LL1 module. Two
of the debris disks, HD 219482 and HD~166, show strong evidence of
24~$\mu$m emission. For comparison, we also show the
spectra of three stars without any significant infrared excess in Figure
\ref{controls}. These spectra tightly hug the line of zero fractional
excess at all wavelengths and do not show a positive fractional excess
slope in LL1. Note that the spectrum of HD 50499 shows the SL1 order
curvature between 10 and 13~$\mu$m. As explained in \S \ref{detections},
we chose the first ten unclipped resolution elements of LL1
(20.93--22.45~$\mu$m) to define our photosphere in order to avoid this
order droop.

Our overall detection rate of debris disks is 11\%, but the non-uniform
sensitivity and the inclusion of 70~$\mu$m follow-up observations in our
survey bias this statistic. Because of the non-uniformity of our
observations, we do not attempt to calculate an error bar for our
detection rate.  However, our results agree with the IRS study of
Beichman et al.\ (2006b), who found 32~$\mu$m excesses around $12 \pm
5$\% of stars in a sample that also included follow-ups of 70~$\mu$m
excesses. Similarly, Lawler et al.\ (2009) report a 32~$\mu$m excess
detection rate of $11.8 \pm 2.4$\%.


The most intrinsically luminous debris disk in our sample is that
orbiting the planet host HD~202206, which has $L_{\rm dust} / L_* >
10^{-4}$.  For comparison, the well-studied debris disks orbiting the
planet hosts HD~69830 and HD~82943 have luminosity ratios $L_{\rm dust}
/ L_*$ of $2 \times 10^{-4}$ and $1.2 \times 10^{-4}$ respectively
(Beichman et al.\ 2005). All but two of the 111 stars in our sample have
a minimum $3 \Sigma$ detectability of $L_{\rm dust} / L_* < 10^{-4}$;
thus only $\sim 1$\% of the sample has a fractional dust luminosity
brighter than $10^{-4} L_*$. Our result agrees with the $2
\pm 2$\% detection rate for excesses of $L_{\rm dust} / L_*$ calculated
by Bryden et al.\ (2006).





\section{Dust Temperature and Location}
\label{dtemp}

We will now determine the location of the debris for the sample of 11
stars with infrared excesses detected at 32~$\mu$m.  Of our 11 detected
debris disks, nine also have excess emission at 70~$\mu$m (Beichman et
al.\ [2006a], Trilling et al.\ [2008], Bryden et al.\ [2009],
K\'{o}sp\'{a}l et al.\ [2009]). Table 2 lists the blackbody temperature
of each debris disk detected at multiple wavelengths.  Beichman et al.\
(2006a) detected HD 219482's dust at both 24 and 70~$\mu$m and
calculated $T_{\rm dust} = 82 \pm 3$~K for large grains, in excellent
agreement with our value of $81 \pm 3$~K. Our $79 \pm 3$~K temperature
for the dust orbiting HD~166 does not agree with Tanner et al.\ (2009),
who calculated a lower limit of 104~K using the previously published
70~$\mu$m detection (Trilling et al.\ 2008) and the 160~$\mu$m upper
limit. However, grain temperature varies with size roughly as
\begin{equation}
T_{\rm dust} = 4.0 \left ( \frac{L_*}{r_{\rm gr}} \right
)^{1/6} a^{-1/3} ,
\label{grainsize}
\end{equation}
where $r_{\rm gr}$ is the grain radius and $a$ is the semimajor axis of
the debris orbit (Krugel 2003). Therefore each photometric band is most
sensitive to a certain combination of grain size and distance from the
star. Different pairs of photometric bands may, therefore, lead to
different grain temperature distance estimates (Carpenter et al.\ 2009).

\begin{figure}
\plotone{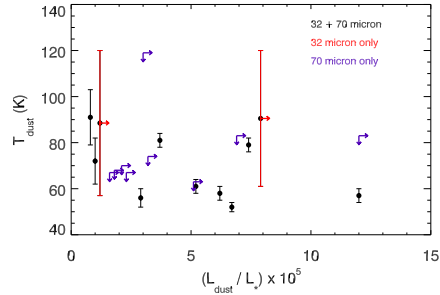}
\caption{Dust luminosity vs. temperature for all debris disks in our
sample. Black circles indicate the debris disks detected at both 32 and
70~$\mu$m. Red error bars indicate the possible temperature range for
debris disks detected only at 32~$\mu$m while right-pointing red arrows
denote that the luminosity ratios for these debris disks are lower
limits only. Purple points with double arrows denote the debris disks
detected only at 70~$\mu$m, which gives upper limits for temperatures
and lower limits for luminosities. Only four stars, HD~142, HD~33262,
HD~108874 and HD~130322, may have debris rings with dust hotter than
100~K (see \S \ref{dtemp}).}
\label{templum}
\end{figure}

By combining our IRS observations with previous MIPS data, we have
obtained temperature estimates for 15 planet hosts and five non-host
stars.  Figure \ref{templum} shows luminosity ratio vs.\ dust
temperature for all debris disks in our sample, whether detected in this
study or previous 70~$\mu$m MIPS work. The warmest temperature of a
debris disk detected at multiple wavelengths is 91~K for the dust around
HD~33262.  Only four stars may possibly host dust warmer than
100~K---HD~142, HD~33262, HD~108874 and HD~130322 have temperature upper
limits of 117~K, 103~K, 120~K and 120~K respectively. Our finding that
debris disks around solar-type stars are primarily cold agrees with
Hillenbrand et al.\ (2008), who find dust temperatures $< 45-85$~K for
the 25 debris disks in the FEPS {\it Spitzer} Legacy Science Program.
70~$\mu$m debris systems with temperature upper limits only may be true
Kuiper Belt analogs, as most have $T_{\rm dust} < 70$~K.

With the exception of three of the upper limits quoted above, all dust
temperatures are below the water ice sublimation temperature of $T \sim
110$~K in a zero-pressure medium. However, in optically thin disks
photodesorption is the dominant volatile removal mechanism and is
efficient well beyond the thermal ice line (Grigorieva et al.\ 2007).
Indeed, the survival timescale for ice deposited in monolayers on the
surfaces of small grains in an optically thin disk is of order a few
minutes (Chen et al.\ 2005).  Assuming, accordingly, that (a) the dust
in each system is rocky or carbonaceous with an albedo near zero and (b)
most grains are larger than 10~$\mu$m and so have absorption
coefficients $Q \sim 1$ for visible and near-IR irradiation, we can find
the distance of the dust from its host star in the blackbody
approximation, appropriate for large grains:
\begin{equation}
a = \frac{R_*}{2} \frac{T_*^2}{T_{\rm dust}^2} ,
\label{dustdist}
\end{equation}
where $R_*$ is the star radius. For our solar-type stars with $60 \la
T_{\rm dust} \la 100$~K, we find $7 {\rm AU} \la a \la 20 {\rm AU}$.
Large grains in the debris disks we have discovered at 32~$\mu$m occupy
what would be the region between Saturn and Uranus in our own Solar
System.

Smaller grains in the 60--100~K temperature range would reside
on larger orbits. We estimate the minimum grain size in our
debris disks from the radiation pressure blowout limit
(Artymowicz 1988) of
\begin{equation}
r_{\rm min} = \frac{3 L_* Q_{\rm pr}}{16 \pi G M_* c \rho},
\label{amin}
\end{equation}
where $r_{\rm min}$ is the minimum grain size, $L_*$ and $M_*$ are the
star luminosity and mass, $Q_{\rm pr}$ is the radiation pressure
coupling coefficient, and $\rho$ is the internal density of the grains.
$Q_{\rm pr} = 1$ is a good approximation for Solar-type stars (Krugel
2003).  For silicate grains with $\rho = 3.3$~g~cm$^{-3}$, $a_{\rm min}
= 0.1 \; \mu$m. Icy grains with $\rho = 1.0$~g~cm$^{-3}$ orbiting a
Solar-type star have minimum size $a_{\rm min} = 0.3 \; \mu$m. Using
equation \ref{grainsize}, which applies to grain sizes from
approximately 0.03 to 30~$\mu$m (Kruegel 2003), we find that 70~K
silicate grains at the blowout limit orbit $\sim 240$~AU from Solar-type
stars. Low-density icy grains, which would be destroyed quickly by
photodesorption, are limited to within $\sim 142$~AU from their host
stars.

In the introduction, we emphasized the lack of any dynamical connection
between short-period planets and cold dust---planets detected by radial
velocity neither migrated through and destroyed our planetesimal belts,
nor dynamically stir them today. Based on the 240~AU maximum orbital
radius of grains with temperatures near 70~K, a representative
temperature for our detected debris disks, we can restrict the
formation locations of the known radial-velocity planets to within $\sim
240$~AU. In any individual debris system where the bulk of the infrared
excess emission comes from large, near-blackbody grains, we can restrict
the giant planet formation zone even further, excluding regions beyond
15~AU. In such systems, giant planet formation by core accretion would
have been restricted to a narrow belt near the ice line, consistent with
the theoretical predictions of Dodson-Robinson et al.\ (2009b). So far,
the orbital radii of the 15 debris disks orbiting planet hosts in this
sample provide the only purely observational constraint on where the
radial-velocity planets could have formed.

\section{Conclusions}
\label{conclusions}

We searched a sample of 111 stars, including 105 planet hosts, for
30-34~$\mu$m excess emission indicating circumstellar dust. Out of 111
stars, 11 showed detectable 32~$\mu$m spectrophotometric excess. Our
debris disk detection rate of 11\% is consistent with previous IRS and
MIPS 70~$\mu$m studies of solar-type stars. We have identified two new
debris disks orbiting HD~108874 and HD~130322 and calculated dust
temperatures and luminosities for the debris disks that had been
previously detected at 70~$\mu$m.

Our IRS data show that the majority of debris systems contain cold dust
with $T_{\rm dust} < 100$~K. Here we confirm earlier results from
24~$\mu$m photometry (Bryden et al.\ 2006) and IRS spectroscopy
(Hillenbrand et al.\ 2008) demonstrating that excess emission is usually
detectable only at wavelengths longer than 30~$\mu$m. The exception in
our survey that has noticeable excess at 20~$\mu$m, HD~166, is cooler
than 80~K. Lower limits to the circumstellar dust temperatures of
Solar-type stars come from the MIPS 160~$\mu$m photometry of Tanner et
al.\ (2009); they find that $T_{\rm dust} > 27$~K for the majority of
the debris disks in their sample.  While all of our debris disks are
colder than the ice sublimation temperature, we expect them to contain
mainly bare rock/carbonaceous grains since photoevaporation of ice is
extremely efficient in optically thin disks (Grigorieva et al.\ 2007).

The cold temperatures of most circumstellar dust disks make them
excellent candidates for follow-up with the Herschel Space Observatory
(Pilbratt et al.\ 2010). The PACS medium-resolution spectrometer, which
covers 55-210~$\mu$m at a resolution of $R = \lambda / \Delta \lambda =
$1000-4000 and also includes photometric bands, will cover the peak
blackbody emission of such disks and provide detailed dust spectral
energy distributions. Circumstellar dust SEDs constrain the location of
the emitting dust and show whether it occupies a narrow ring---the
simplifying assumption used in this work---or a broad, multi-AU band
(e.g.\ Hillenbrand et al.\ 2008). If any debris disks contain icy dust
produced by very recent collisions, PACS may be able to detect H$_2$O
gas produced by photoevaporation of the grain mantles.

One question open for debate is why the planetesimal belts surrounding
our host stars did not accumulate into giant planets. While planets may
shape debris disk edges (Quillen 2006), they will eject planetesimals
within a few Roche lobe radii of their orbits (Bryden et al.\ 2000), so
planets and circumstellar dust should not coincide. Planet-formation
simulations show that formation of ice giants such as Uranus and Neptune
is robust out to $\sim 15$~AU in disks of several times the minimum-mass
solar nebula (Dodson-Robinson et al.\ [2009a], Dodson-Robinson \&
Bodenheimer [2010]). Blackbody grains with $T_{\rm dust} = 70$~K reside
only 15~AU from solar-type stars.

If most of our debris disk luminosity comes from blackbody grains, a
simple explanation for the presence of planetesimal belts in locations
that could theoretically host planets is that the majority of
protostellar disks do not have enough mass to form ice giants near $\sim
15$~AU. Gas and ice giant formation by core accretion would then be
restricted to a narrow region around the ice line in most disks, as
predicted by Dodson-Robinson et al.\ (2009b). The Sun's ability to form
Uranus and Neptune in the trans-Saturnian region (Dodson-Robinson \&
Bodenheimer 2010) would be the unusual result of an abnormally massive
solar nebula.

If most of the infrared excess emission is produced by small grains near
the radiation-pressure blowout limit, we can in general confine the
ensemble of radial-velocity planets to formation zones less than 240~AU
from host stars. We have therefore presented the first fully
observational constraints on the formation locations of the ensemble of
radial-velocity planets.


Support for S.D.R.'s work at NASA Exoplanet Science Institute was
provided by NASA through the {\it Spitzer} Space Telescope Fellowship
Program. S.D.R.'s work at University of Texas was supported by the
Dean's Fellowship program of UT's College of Natural Sciences. J.C. was
partially supported by a contract from JPL/Caltech. S.D.R. acknowledges
input on IRS data reduction from Joel Green. This research has made use
of the following online resources: the SIMBAD database, VizieR catalogue
access tool, and Aladin sky atlas operated at CDS, Strasbourg, France;
the NASA/IPAC/NExScI Star and Exoplanet Database and the NASA/IPAC
Infrared Science Archive, which are operated by the Jet Propulsion
Laboratory, California Institute of Technology, under contract with the
National Aeronautics and Space Administration; and the Extrasolar
Planets Encyclopaedia at http://exoplanet.eu.

\clearpage
\LongTables
\begin{landscape}
\begin{deluxetable*}{rllccccccc}
\tabletypesize{\tiny}
\tablewidth{0pt}
\tablecaption{Observed Stellar Sample}
\tablehead{
\colhead{HD} & \colhead{Name} & \colhead{Spectral Type} &
\colhead{$T_{\rm eff} (K)$} &
\colhead{$T_{\rm eff}$ Reference} & \colhead{$V$ (mag)} &
\colhead{Dist. (pc)} &
\colhead{Planets} & 
\colhead{$ {L_{\rm dust} \over L_*} \times 10^5 $ (70~$\mu$m)} &
\colhead{70~$\mu$m Reference}
}
\startdata
\nodata & GJ 581         & M3             & 3310 & Bessell (1995) & 10.57 & 6.3 & 3 & 8.49 & Kospal et al.\ (2009) \\
\nodata & GJ 876      & M4              & 3130 & Bessell (1995) & 10.17 & 4.7 & 3 & $<$3.2 & Kospal et al.\ (2009) \\
\nodata & Hip 14810       & G5            & 5485 & Wright et al.\ (2009) & 8.52 & 52.9 & 3 & \nodata & \nodata \\
142 & & F7V             & 6248 & VF05\tablenotemark{a} & 5.70 & 25.6 & 1 & 0.8 & B09\tablenotemark{b} \\
166 & & K0V & 5577 & VF05 & 6.13 & 13.7 & 0 & 6.0 & Trilling et al.\ (2008) \\
1237 & GJ 3021         & G8.5Vk: & 5360 & Santos et al.\ (2004b) & 6.59 & 17.6 & 1 & $<$1.0 & B09 \\
3651 & & K0V          &  5220 & VF05 & 5.80 & 11.1 & 1 &  $<$1.1& B09 \\
4203 & & G5         & 5701 & VF05 & 8.69 & 77.8 & 1 &  $<$23.9 & B09 \\
4208 & & G7VFe-1H-05  & 5600 & VF05 & 7.79 & 32.7 & 1 &  $<$2.9 & B09 \\
8574 & & F8  & 6049 & VF05 & 7.11 & 44.2 & 1 & $<$1.5 & B09\\
10697 & 109 Psc & G5IV    & 5680 & VF05 & 6.29 & 32.6 & 1 &  $<$1.3 & B09 \\
11964 & & G5              & 5349 & VF05 & 6.42 & 34.0 & 2 &  $<$1.0 & B09 \\
12661 & & K0V             & 5742 & VF05 & 7.44 & 37.2 & 2 &  $<$6.5 & B09 \\
13445 & GJ 86        & K1V   & 5150 & VF05 & 6.17 & 10.9 & 1  & \nodata & \nodata \\
16141 & & G5IV            & 5793 & VF05 & 6.78 & 35.9 & 1 &  $<$2.6 & B09 \\
17051 & HR 810  & F9VFe+03  & 6038 & Rocha-Pinto \& Maciel (1998) & 5.40 & 17.2 & 1 &  $<$0.5 & B09 \\
19994 & & F8V             & 6188 & VF05 & 5.06 & 22.4 & 1 &  0.5 & B09  \\
20367 & & G0              & 6100 & Santos et al.\ (2004b) & 6.41 & 27.1 & 1 &  $<$1.1 & B09 \\
20782 & & G1.5V           & 5758 & VF05 & 7.38 & 36.0 & 1 &  $<$2.1 & B09 \\
23079 & & F9.5V           & 5927 & VF05 & 7.12 & 34.6 & 1 &  $<$0.9 & B09 \\
23127 & & G2V             & 5752 & VF05 & 8.58 & 89.1 & 1 &  $<$7.3 & B09 \\
23596 & & F8    & 5903 & VF05 & 7.25 & 52.0 & 1 &  $<$3.1 & B09 \\
27442 & $\epsilon$~Ret & K2III   & 4845 & VF05 & 4.44 & 18.2 & 1 &  $<$0.3 & B09 \\
27894 & & K2V             & 4875 & Santos et al.\ (2005) & 9.42 & 42.4 &  1 & $<$10.6 & B09 \\
28185 & & G6.5IV-V        & 5656 & Santos et al.\ (2004b) & 7.80 & 39.6 &  1 & $<$5.7 & B09 \\
30177 & & G8V  & 5607 & VF05 & 8.41 & 54.7 & 1 &  $<$11.2 & B09 \\
33262 & $\zeta$ Dor & F7V & 6200 & Flower (1996) & 4.72 & 11.7 & 0 &  0.51 & Trilling et al.\ (2008) \\
33283 & & G4V & 5995 & Johnson et al.\ (2006) & 8.05 & 86.9 & 1 &  $<$4.9 & B09 \\
33564 & & F6V  & 6250 & Nordstrom et al.\ (2004) & 5.10 & 21.0 & 1  & $<$0.5 & B09  \\
33636 & & G0VH-03 & 5904 & VF05 & 7.06 & 28.7 & 0  & 5.1 & Trilling et al.\ (2008) \\
37124 & & G4IV-V   & 5500 & VF05 & 7.68 & 33.2 & 3 &  $<$7.2 & B09 \\
37605 & & K0   & 5475 & Cochran et al.\ (2004) & 8.69 & 42.9 & 1 &  $<$23.2 & B09 \\
40979 & & F8  & 6089 & VF05 & 6.73 & 33.3 & 1 &  1.44 & Kospal et al.\ (2009) \\
41004 & HD 41004 A & K1V & 5010 & Santos et al.\ (2002) & 8.65 & 43.0 & 1 &  $<$6.9 & B09 \\
45350 & & G5 & 5616 & VF05 & 7.88 & 48.9 & 1 &  $<$7.2 & B09 \\
46375 & & K1IV & 5285 & VF05 & 7.84 & 33.4 & 1 &  17.8 & Kospal et al.\ (2009) \\
49674 & & G0 & 5662 & VF05 & 8.10 & 40.7 & 1 &  $<$4.5 & B09 \\
50499 & & G1V  & 6069 & VF05 & 7.21 & 47.3 & 1 &  1.6 & Kospal et al.\ (2009) \\
50554 & & F8V & 5928 & VF05 & 6.84 & 31.0 & 1 &  5.85 & Trilling et al.\ (2008) \\
52265 & & G0V  & 6076 & VF05 & 6.30 & 28.1 & 1 &  2.69 & Trilling et al.\ (2008) \\
63454 & & K3V & 4841 & VF05 & 9.37 & 35.8 & 1 &  $<$16.0 & B09 \\
65216 & & G5V  & 5666 & Santos et al.\ (2003) & 7.97 & 35.6 & 1 &  $<$4.4 & B09 \\
68988 & & G0  & 5960 & VF05 & 8.20 & 58.8 & 1 &  $<$4.2 & B09 \\
70642 & & G6VCN+05 & 5705 & VF05 & 7.18 & 28.8 & 1 &  $<$6.3 & B09\\
72659 & & G0  & 5919 & VF05 & 7.46 & 51.4 & 1 &  $<$2.2 & B09 \\
73256 & & G8IV-VFe+05 & 5570 & Udry et al.\ (2003) & 8.08 & 36.5 & 1 &  $<$5.5 & B09 \\
74156 & & G0 & 6067 & VF05 & 7.61 & 64.6 & 3 &  $<$3.4 & B09 \\
75289 & & F9VFe+03  & 6095 & VF05 & 6.36 & 28.9 & 1 &  $<$14.9 & B09 \\
76700 & & G6V & 5668 & VF05 & 8.13 & 59.7 & 1 &  $<$3.7 & B09 \\
81040 & & G0  & 5700 & Sozzetti et al.\ (2006) & 7.72 & 32.6 & 1 &  $<$3.2 & B09\\
83443 & & K0V & 5453 & VF05 & 8.24 & 43.5 & 1 &  $<$9.0 & B09 \\
88133 & & G5IV & 5494 & Fischer et al.\ (2005) & 8.06 & 74.5 & 1 &  \nodata & \nodata \\
89307 & & G0V & 5897 & VF05 & 7.01 & 30.9 & 1 &  $<$2.8 & B09 \\
89744 & & F7V & 6291 & VF05 & 5.74 & 39.0 & 1 &  $<$0.9 & B09 \\
93083 & & K2IV-V & 4995 & Lovis et al.\ (2005) & 8.33 & 28.9 & 1 &  $<$9.2 & B09 \\
99492 & & K2V  & 4954 & VF05 & 7.57 & 18.0 & 1 &  $<$5.1 & B09 \\
102117 & & G6V  & 5695 & VF05 & 7.47 & 42.0 & 1 &  $<$16.8 & B09\\
102195 & & K0 & 5291 & Melo et al.\ (2007) & 8.06 & 29.0 & 1 &  $<$9.2 & B09 \\
105211 & $\eta$ Cru & F2V & 6600\tablenotemark{c} & Peletier (1989) & 4.15 & 19.7 & 0 &  6.9 & Beichman et al.\ (2006a) \\
107148 & & G5   & 5797 & VF05 & 8.01 & 51.3 & 1 &  $<$5.6 & B09 \\
108147 & & F8VH+04 & 6156 & VF05 & 7.00 & 38.6 & 1 &  $<$23.9 & B09 \\
108874 & & G5 & 5550 & VF05 & 8.74 & 68.5 & 2 &  $<$13.6 & B09 \\
109749 & & G3V & 5903 & Fischer et al.\ (2006) & 8.08 & 59.0 & 1 &  $<$6.2 & B09 \\
111232 & & G8VFe-10 & 5494 & Santos et al.\ (2004b) & 7.61 & 28.9 & 1 &  $<$6.1 & B09 \\
114386 & & K3V & 4819 & VF05 & 8.73 & 28.0 & 1 &  $<$6.8 & B09 \\
114762 & & F9V  & 5952 & VF05 & 7.30 & 40.6 & 1 &  $<$3.3 & B09 \\
114783 & & K0 & 5135 & VF05 & 7.56 & 20.4 & 1 &  $<$3.0 & B09\\
117176 & 70 Vir & G5V & 5544 & VF05 & 5.00 & 18.1 & 1 &  0.97 & Trilling et al.\ (2008) \\
117207 & & G7IV-V  & 5723 & VF05 & 7.26 & 33.0 & 1 &  $<$2.1 & B09 \\
117618 & & G0V & 5963 & VF05 & 7.17 & 38.0 & 1 &  $<$2.3 & B09 \\
118203 & & K0 & 5695 & da Silva et al.\ (2006) & 8.05 & 88.6 & 1 &  $<$13.2 & B09\\
130322 & & K0III & 5308 & VF05 & 8.04 & 29.8 & 1 &  $<$7.9 & B09\\
134987 & 23 Lib & G5V & 5750 & VF05 & 6.45 & 25.7 & 2 &  $<$2.70 & Trilling et al.\ (2008) \\
136118 & & F8 & 6097 & VF05 & 6.93 & 52.3 & 1 &  $<$1.3 & B09 \\
142415 & & G1V & 5901 & VF05 & 7.34 & 34.6 & 1  & $<$10.9 & B09 \\
143761 & $\rho$~Cr B & GOV & 5823 & VF05 & 5.40 & 17.4 & 1 &  $<$0.6 & B09 \\
145675 & 14 Her & K0V & 5347 & VF05 & 6.67 & 18.1 & 2 &  $<$1.0 & B09 \\
147513 & & G1VH-04 & 5929 & VF05 & 5.38 & 12.9 & 1 &  $<$1.6 & B09 \\
149026 & & G0IV & 6147 & Sato et al.\ (2005) & 8.15 & 78.9 & 1 &  $<$5.2 & B09 \\
149143 & & G0 & 5884 & Fischer et al.\ (2006) & 7.89 & 63.5 & 1 &  $<$4.6 & B09\\
154857 & & G5V & 5605 & VF05 & 7.24 & 68.5 & 1 &  $<$4.7 & B09 \\
159868 & & G5V & 5623 & VF05 & 7.24 & 52.7 & 1 &  $<$9.9 & B09 \\
160691 & $\mu$ Arae & G3IV-V & 5784 & Santos et al.\ (2004a) & 5.15 & 15.3 & 4 &  $<$0.9 & B09 \\
162020 & & K2V & 4844 & VF05 & 9.10 & 31.3 & 1 &  $<$28.6 & B09\\
168443 & & G6V & 5579 & VF05 & 6.92 & 37.9 & 2 &  $<$14.7 & B09 \\
169830 & & F7V & 6221 & VF05 & 5.91 & 36.3 & 2 &  $<$1.5 & B09 \\
177830 & & K0 & 4948 & VF05 & 7.18 & 59.0 & 1 &  $<$1.4 & B09 \\
183263 & & G2IV & 5936 & VF05 & 7.86 & 52.8 & 2 &  $<$9.1 & B09 \\
185269 & & G0IV & 5980 & Johnson et al.\ (2006) & 6.68 & 47.4 & 1 &  $<$2.4 & B09 \\
186427 & 16 Cyg B & G3V & 5674 & VF05 & 6.20 & 21.4 & 1 &  $<$1.3 & B09 \\
187085 & & G0V & 6075 & VF05 & 7.22 & 45.0 & 1 &  2.3 & Kospal et al.\ (2009) \\
187123 & & G5 & 5814 & VF05 & 7.83 & 47.9 & 2 &  $<$10.6 & B09 \\
188015 & & G5IV & 5745 & VF05 & 8.24 & 52.6 & 1 &  $<$29.4 & B09 \\
188753 & HD 188753 A & G8V & 5750 & Konacki (2005) & 7.40 & 44.8 & 0\tablenotemark{d} &  $<$8.7 & B09 \\
190360 & & G7IV-V & 5551 & VF05 & 5.71 & 15.9 & 2 &  $<$9.7 & B09\\
192263 & & K2V & 4975 & VF05 & 7.79 & 19.9 & 1 &  5.4 & B09 \\
192699 & & G5 & 5220 & Johnson et al.\ (2007) & 6.45 & 67.4 & 1 &  $<$1.1 & B09 \\
195019 & & G3IV-V & 5788 & VF05 & 6.91 & 37.4 & 1 &  $<$2.1 & B09  \\
196050 & & G3V & 5892 & VF05 & 7.50 & 46.9 & 1 &  $<$2.7 & B09 \\
196885 & & F8IV & 6185 & VF05 & 6.40 & 33.0 & 1 &  $<$1.0 & B09 \\
202206 & & G6V & 5787 & VF05 & 8.08 & 46.3 & 2 &  14.3 & B09 \\
208487 & & G2V & 6067 & VF05 & 7.47 & 44.0 & 1 &  $<$4.9 & B09 \\
210277 & & G0V  & 5555 & VF05 & 6.63 & 21.3 & 1 &  $<$0.8 & B09 \\
212301 & & F8V & 6256 & Lo Curto et al.\ (2006) & 7.76 & 52.7 & 1 &  $<$4.9 & B09 \\
213240 & & G0/G1V & 5967 & VF05 & 6.80 & 40.7 & 1 &  $<$1.1 & B09 \\
216435 & & G0V & 5999 & VF05 & 6.03 & 33.3 & 1 &  2.2 & B09 \\
216770 & & G9VCN+1 & 5229 & Mayor et al.\ (2004) & 8.10 & 37.9 & 1 &  $<$19.0 & B09\\
217107 & & G8IV & 5704 & VF05 & 6.18 & 19.7 & 2 &  $<$1.5 & B09 \\
219482 & & F6V & 6240 & Flower (1996) & 5.66 & 20.6 & 0 &  2.8 & Beichman et al.\ (2006a) \\
222404 & $\gamma$~Cep & K1IV & 4916 & Santos et al.\ (2004b) & 3.23 & 13.8 & 1 &  $<$0.3 & B09 \\
222582 & & G5 & 5726 & VF05 & 7.68 & 41.9 & 1 &  $<$2.3 & B09 \\
224693 & & G2V & 6037 & Butler et al.\ (2006) & 8.23 & 94.1 & 1 &  $<$8.0 & B09 \\
\enddata
\tablenotetext{a}{VF05 = Valenti \& Fischer (2005)}
\tablenotetext{b}{B09 = Bryden et al.\ (2009)}
\tablenotetext{c}{Based on tabulated values for $T_{\rm eff}$
vs.\ $V-K$}
\tablenotetext{d}{Planet's existence under debate;
see Eggenberger et al.\ (2007b)}
\end{deluxetable*}
\clearpage
\end{landscape}

\begin{deluxetable*}{rcrrrrrr}
\tabletypesize{\scriptsize}
\tablewidth{0pt}
\tablecaption{IRS 30-34~$\mu$m Spectrophotometric Results}
\tablehead{
\colhead{HD} & \colhead{Name} & \colhead{$F_*$} &
\colhead{$F_{\rm obs}$} & \colhead{$\Sigma$} &
\colhead{$F_{\rm dust} / F_* $} &
\colhead{${L_{\rm dust} \over L_*} \times
10^5$}\tablenotemark{a} &
\colhead{$T_{\rm dust}$ (K)}\tablenotemark{b}
}
\startdata
\nodata & GJ 581  & 16.1 & 17.1 & 0.99 & 0.06 & $> 12$ & $< 83$ \\
\nodata & GJ 876  & 24.8 & 25.3 & 0.67 & 0.02 & $< 5.5$ & \\
\nodata & Hip 14810 & 7.5 & 7.0 & 0.93 & -0.07 & $< 3.7$ & \\
142 & & 30.9 & 30.8 & 2.67 & 0.00 & $> 3.0$ & $< 119$ \\
{\bf 166} & & 78.9 & 119 & 2.00 & 0.50 & 7.4 & $79 \pm 3$ \\
1237 & GJ 3021 & 39.9 & 38.7 & 1.71 & -0.03 & $< 1.7$ & \\
3651 & & 98.6 & 101.0 & 2.29 & 0.02 & $< 0.8$ & \\
4203 & & 5.8 & 5.9 & 0.40 & 0.01 & $< 2.0$ & \\
4208 & & 12.8 & 13.1 & 0.87 & 0.02 & $< 2.0$ & \\
8574 & & 19.2 & 17.7 & 1.57 & -0.08 & $< 1.9$ & \\
10697 & 109 Psc & 48.7 & 48.1 & 1.98 & -0.01 & $< 1.2$ & \\
11964 & & 8.3 & 8.3 & 1.39 & 0.00 & $< 5.7$ & \\
12661 & & 18.7 & 19.6 & 0.86 & 0.05 & $< 1.4$ & \\
13445 & GJ 86 & 90.3 & 88.3 & 3.00 & -0.02 & $< 1.3$ & \\
16141 & & 32.4 & 31.4 & 1.72 & -0.03 & $< 1.4$ & \\
17051 & HR 810 & 107.4 & 108.9 & 2.34 & 0.01 & $< 0.5$ & \\
19994 & & 126.0 & 124.0 & 2.91 & -0.02 & $> 6.9$ & $< 83$ \\
20367 & & 76.0 & 76.1 & 1.43 & 0.00 & $< 0.4$ & \\
20782 & & 20.3 & 20.0 & 1.29 & -0.02 & $< 1.7$ & \\
23079 & & 20.1 & 21.0 & 1.11 & 0.05 & $< 1.3$ & \\
23127 & & 5.5 & 4.9 & 0.65 & -0.10 & $< 2.8$ & \\
23596 & & 24.9 & 25.0 & 1.16 & 0.00 & $< 1.1$ & \\
27442 & $\epsilon$~Ret & 241.0 & 240.9 & 3.52 & -0.30 & $< 0.4$ & \\
27894 & & 2.5 & 2.5 & 0.67 & -0.02 & $< 12$ & \\
28185 & & 13.2 & 13.1 & 0.59 & -0.01 & $< 1.3$ & \\
30177 & & 8.4 & 8.0 & 0.51 & -0.05 & $< 1.8$ & \\
{\bf 33262} & $\zeta$~Dor & 177.3 & 192.9 & 1.64 & 0.09 & 0.8 & $91 \pm 12$ \\
33283 & & 8.5 & 9.5 & 0.81 & 0.12 & $< 2.2$ & \\
33564 & & 109.6 & 112.3 & 2.78 & 0.03 & $< 0.5$ & \\
{\bf 33636} & & 24.4 & 29.7 & 0.85 & 0.22 & 5.2 & $61 \pm 3$ \\
37124 & & 16.4 & 14.9 & 1.18 & -0.09 & $< 2.2$ & \\
37605 & & 7.6 & 8.4 & 0.51 & 0.11 & $< 2.3$ & \\
40979 & & 27.4 & 28.7 & 0.90 & 0.05 & $> 1.6$ & $< 67$ \\
41004 & HD 41004 A & 11.7 & 12.2 & 0.61 & 0.04 & $< 2.5$ & \\
45350 & & 6.4 & 6.4 & 0.47 & -0.01 & $< 2.2$ & \\
46375 & & 10.7 & 11.5 & 6.66 & 0.07 & $> 3.2$ & $< 74$ \\
49674 & & 10.3 & 10.6 & 0.40 & 0.03 & $< 1.2$ & \\
50499 & & 18.1 & 18.3 & 0.65 & 0.01 & $> 1.8$ & $< 68$ \\
{\bf 50554} & & 25.6 & 31.0 & 0.68 & 0.21 & 6.2 & $58 \pm 3$ \\
{\bf 52265} & & 41.0 & 44.5 & 0.66 & 0.09 & 2.9 & $56 \pm 4$ \\
63454 & & 7.1 & 7.3 & 0.47 & 0.03 & $< 3.5$ & \\
65216 & & 11.4 & 10.1 & 0.63 & -0.11 & $< 1.6$ & \\
68988 & & 8.3 & 8.9 & 0.32 & 0.08 & $< 1.0$ & \\
70642 & & 23.5 & 23.0 & 0.63 & -0.02 & $< 0.8$ & \\
72659 & & 16.5 & 17.7 & 0.67 & 0.08 & $< 1.0$ & \\
73256 & & 12.3 & 11.8 & 0.52 & -0.05 & $< 1.4$ & \\
74156 & & 12.2 & 11.5 & 0.58 & -0.06 & $< 1.1$ & \\
75289 & & 38.1 & 31.7 & 1.54 & -0.17 & $< 0.9$ & \\
76700 & & 10.4 & 10.0 & 0.70 & -0.05 & $< 2.0$ & \\
81040 & & 16.0 & 16.5 & 0.84 & 0.03 & $< 1.6$ & \\
83443 & & 10.8 & 9.5 & 0.41 & -0.12 & $< 1.3$ & \\
88133 & & 14.7 & 14.0 & 0.56 & -0.05 & $< 1.3$ & \\
89307 & & 25.5 & 25.0 & 0.84 & -0.02 & $< 0.9$ & \\
89744 & & 53.7 & 53.8 & 2.47 & 0.00 & $< 1.0$ & \\
93083 & & 10.0 & 10.0 & 0.46 & 0.00 & $< 2.1$ & \\
99492 & & 30.6 & 31.4 & 1.08 & 0.03 & $< 0.8$ & \\
102117 & & 19.5 & 20.4 & 1.09 & 0.05 & $< 1.7$ & \\
102195 & & 13.3 & 13.9 & 0.42 & 0.04 & $< 1.1$ & \\
{\bf 105211} & $\eta$~Cru & 178.5 & 215.6 & 3.30 & 0.21 & 6.7 & $51 \pm 2$ \\
107148 & & 10.4 & 9.6 & 0.68 & -0.08 & $< 1.7$ & \\
108147 & & 29.4 & 29.2 & 0.99 & -0.01 & $< 0.8$ & \\
{\bf 108874} & & 5.9 & 7.9 & 0.33 & 0.32 & $> 1.2$ & $57 < T < 120$ \\
109749 & & 8.0 & 7.5 & 0.61 & -0.06 & $< 1.8$ & \\
111232 & & 54.4 & 54.3 & 1.26 & 0.00 & $< 0.7$ & \\
114386 & & 12.5 & 13.7 & 0.71 & 0.09 & $< 3.0$ & \\
114762 & & 19.3 & 19.7 & 1.13 & 0.02 & $< 1.5$ & \\
114783 & & 21.1 & 20.7 & 1.39 & -0.02 & $< 2.6$ & \\
{\bf 117176} & 70 Vir & 173.0 & 182.1 & 2.87 & 0.05 & 1.0 & $72 \pm 10$ \\
117207 & & 21.1 & 18.6 & 1.61 & -0.12 & $< 2.3$ & \\
117618 & & 21.2 & 20.4 & 1.04 & -0.04 & $< 1.3$ & \\
{\bf 130322} & & 13.0 & 16.1 & 0.58 & 0.24 & $> 7.9$ & $61 < T < 120$ \\
134987 & 23 Lib & 39.9 & 40.4 & 0.76 & 0.01 & $< 0.5$ & \\
136118 & & 26.7 & 26.3 & 0.88 & -0.01 & $< 0.8$ & \\
142415 & & 17.6 & 19.2 & 0.59 & 0.09 & $< 0.9$ & \\
143761 & $\rho$~Cr B & 87.2 & 87.6 & 2.19 & 0.01 & $< 0.6$ & \\
145675 & 14 Her & 53.2 & 50.7 & 2.57 & -0.05 & $< 1.9$ & \\
147513 & & 110.5 & 115.4 & 2.78 & 0.05 & $< 0.7$ & \\
149026 & & 7.8 & 7.9 & 0.35 & 0.02 & $< 1.0$ & \\
149143 & & 10.8 & 11.5 & 0.31 & 0.06 & $< 0.8$ & \\
154857 & & 29.1 & 29.5 & 0.94 & 0.01 & $< 1.0$ & \\
159868 & & 23.71 & 21.93 & 0.78 & -0.075 & $< 0.6$ & \\
160691 & $\mu$~Arae & 247.1 & 247.3 & 2.36 & 0.00 & $< 0.3$ & \\
162020 & & 10.2 & 9.5 & 0.36 & -0.07 & $< 1.9$ & \\
168443 & & 30.4 & 29.8 & 1.70 & -0.02 & $< 1.7$ & \\
169830 & & 50.2 & 51.3 & 1.31 & 0.02 & $< 0.5$ & \\
177830 & & 43.4 & 42.4 & 2.48 & -0.02 & $< 2.6$ & \\
183263 & & 9.7 & 10.8 & 0.51 & 0.12 & $< 1.2$ & \\
185269 & & 34.5 & 35.0 & 1.43 & 0.02 & $< 1.0$ & \\
186427 & 16 Cyg B & 52.6 & 54.2 & 3.74 & 0.03 & $< 2.1$ & \\
187085 & & 16.7 & 16.9 & 0.80 & 0.01 & $> 2.3$ & $< 67$ \\
187123 & & 12.5 & 12.8 & 0.57 & 0.03 & $< 1.2$ & \\
188015 & & 9.1 & 7.5 & 0.91 & -0.18 & $< 3.0$ & \\
188753 & HD 188753 A & 24.6 & 24.3 & 1.17 & -0.01 & $< 1.6$ & \\
190360 & & 91.1 & 94.4 & 1.83 & 0.05 & $< 0.6$ & \\
192263 & & 22.4 & 21.9 & 1.04 & -0.02 & $> 5.1$ & $< 63$ \\
192699 & & 84.0 & 85.6 & 2.48 & 0.02 & $< 1.2$ & \\
195019 & & 31.2 & 32.7 & 0.80 & 0.03 & $< 0.8$ & \\
196050 & & 14.9 & 16.4 & 1.52 & 0.10 & $< 2.7$ & \\
196885 & & 25.9 & 1.00 & 0.00 & $< 0.9$ & \\
{\bf 202206} & & 10.4 & 14.3 & 0.35 & 0.38 & 12 & $57 \pm 3$ \\
208487 & & 13.5 & 12.3 & 0.69 & -0.09 & $< 1.2$ & \\
210277 & & 53.9 & 55.3 & 1.98 & 0.03 & $< 1.3$ & \\
212301 & & 11.2 & 11.7 & 0.64 & 0.04 & $< 1.3$ & \\
213240 & & 34.2 & 34.4 & 1.50 & 0.01 & $< 1.2$ & \\
216435 & & 59.3 & 67.0 & 2.81 & 0.13 & $> 2.1$ & $< 70$ \\
216770 & & 10.2 & 10.4 & 0.55 & 0.02 & $< 1.9$ & \\
217107 & & 56.9 & 55.1 & 3.64 & -0.03 & $< 1.9$ & \\
{\bf 219482} & & 66.5 & 91.0 & 1.15 & 0.37 & 3.7 & $81 \pm 3$\tablenotemark{c} \\
222404 & $\gamma$~Cep & 1622.0 & 1637.0 & 8.36 & 0.01 & $< 0.2$ & \\
222582 & & 12.4 & 12.1 & 0.59 & -0.03 & $< 1.3$ & \\
224693 & & 6.3 & 6.1 & 0.62 & -0.03 & $< 2.3$ & \\
\enddata
\tablenotetext{a}{Upper limits to the dust luminosity are calculated
assuming a single-temperature blackbody with $T_{\rm dust} = 115$~K. For
cold dust undetected by both MIPS and IRS, the true luminosity ratio may
be higher than the value quoted here. See \S \ref{debrisdisks}
for further discussion of the relationship between observed
30-34~$\mu$m fluxes and grain luminosity.}
\tablenotetext{b}{Grain temperature estimates are based on the
blackbody approximation. For a discussion of temperature as a
function of grain size, see \S \ref{dtemp}.}
\tablenotetext{c}{Our temperature measurement for the
circumstellar dust of HD 219482 agrees with that of Beichman et
al.\ (2006), who calculated $T_{\rm dust} = 82 \pm 3$~K based on
24 and 70~$\mu$m photometry.}
\end{deluxetable*}

\end{document}